\documentclass[twocolumn,prl,superscriptaddress,10pt]{revtex4}

\usepackage[usenames]{color}
\usepackage{graphicx}
\usepackage{amsmath}
\usepackage[normalem]{ulem}
\usepackage{amssymb}
\usepackage{bm}
\usepackage{epstopdf}
\usepackage{xcolor}
\usepackage{bbold}
\usepackage{mathrsfs}
\usepackage[section]{placeins}
\usepackage[multiple]{footmisc}

\usepackage{color}
\definecolor{red}{rgb}{0.75,0,0}
\definecolor{blue}{rgb}{0,0,0.75}
\definecolor{green}{rgb}{0,0.5,0}

\newcommand{\microns}{~$\mu$m}

\DeclareMathOperator{\tr}{tr}

\usepackage{verbatim}

\begin{document}


\title{Chiral edge currents in confined fibrosarcoma cells}

\author{V. Yashunsky}
\thanks{these authors contributed equally}
\affiliation{Laboratoire Physico-Chimie Curie, Institut Curie, PSL Research University—Sorbonne Universit\'e, CNRS—Equipe labellis\'ee Ligue Contre le Cancer, 75005 Paris, France}
\email{victoryashunsky@gmail.com}

\author{D. J. G. Pearce}
\thanks{these authors contributed equally}
\affiliation{Department of Physics, University of Geneva, 1211 Geneva, Switzerland}
 \affiliation{Department of Biochemistry, University of Geneva, 1211 Geneva, Switzerland}

\author{C. Blanch-Mercader}
\thanks{these authors contributed equally}
\affiliation{Department of Physics, University of Geneva, 1211 Geneva, Switzerland}
 \affiliation{Department of Biochemistry, University of Geneva, 1211 Geneva, Switzerland}
 
\author{F. Ascione}
\affiliation{Laboratoire Physico-Chimie Curie, Institut Curie, PSL Research University—Sorbonne Universit\'e, CNRS—Equipe labellis\'ee Ligue Contre le Cancer, 75005 Paris, France}

\author{L. Giomi}
\affiliation{Instituut-Lorentz, Universiteit Leiden, P.O. Box 9506, 2300 RA Leiden, The Netherlands}

\author{P. Silberzan}
\affiliation{Laboratoire Physico-Chimie Curie, Institut Curie, PSL Research University—Sorbonne Universit\'e, CNRS—Equipe labellis\'ee Ligue Contre le Cancer, 75005 Paris, France}

\date{\today }


\maketitle



\textbf{During metastatic dissemination, streams of cells collectively migrate through a network of narrow channels within the extracellular matrix \cite{Cheung2016}, before entering into the blood stream. This strategy is believed to outperform other migration modes, based on the observation that individual cancer cells can take advantage of confinement to switch to an adhesion-independent form of locomotion \cite{Paluch2016}. Yet, the physical origin of this behaviour has remained elusive and the mechanisms behind the emergence of coherent flows in populations of invading cells under confinement are presently unknown. Here we demonstrate that human fibrosarcoma cells (HT1080) confined in narrow stripe-shaped regions undergo collective migration by virtue of a novel type of topological edge currents, resulting from the interplay between liquid crystalline (nematic) order, microscopic chirality and topological defects. Thanks to a combination of in vitro experiments and theory of active hydrodynamics \cite{Marchetti2013,Doostmohammadi2018}, we show that, while heterogeneous and chaotic in the bulk of the channel, the spontaneous flow arising in confined populations of HT1080 cells is rectified along the edges, leading to long-ranged collective cell migration, with broken chiral symmetry. These edge currents are fuelled by layers of +1/2 topological defects, orthogonally anchored at the channel walls and acting as local sources of chiral active stress \cite{Duclos2018,Hoffmann2020}. Our work highlights the profound correlation between confinement and collective migration in multicellular systems and suggests a possible mechanism for the emergence of directed motion in metastatic cancer.}

Physiological and pathological conditions, like embryonic morphogenesis, wound healing and cancer invasion, depend upon the ability of cells to migrate collectively \cite{Friedl2009,Ladoux2017,Kawaguchi2017,Hakim2017,Trepat2018}. While commonly ascribed to the cell signaling machinery, it has recently become obvious that such a process further relies on mechanical cues at length scales several times larger than that of individual cells. Depending on the cell shape and motility, as well as the mechanical and geometrical properties of the environment, eukaryotic cell layers have been observed to display a large variety of dynamical behaviors, ranging from collective jamming to chaotic flows reminescent of turbulence in Newtonian fluids \cite{Angelini2011,Park2015,Garcia2015,Atia2018,Blanch-Mercader2018m,Palamidessi2019}. Yet, harnessing collective motion in order to achieve biological functionality, requires a toolbox of reliable and robust control mechanisms, whose exploration and understanding are still in their infancy.


\begin{figure*}[t]
\centering
\includegraphics[width=0.8\textwidth]{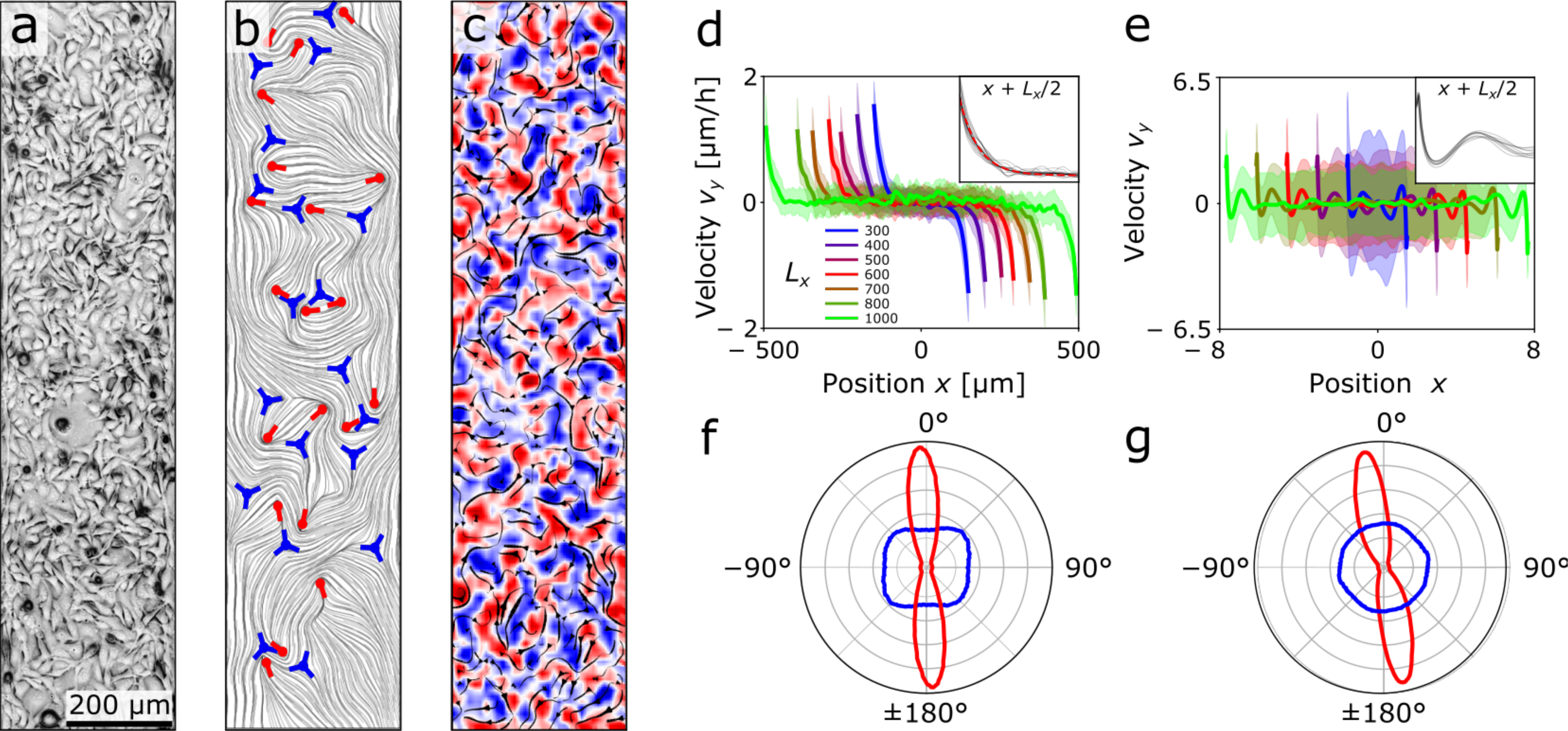}
\caption{\label{fig:F1} HT1080 cell monolayer on an adhesive stripe. (a) Phase contrast image in inverted grayscale. (b) The nematic director field extracted from panel (a) with positive (red) and negative (blue) locations and orientations of half integer defects. (c) Black velocity streamlines are overlaid on the normalised vorticity field (blue-clockwise, red-counter-clockwise). (d) Average flow along the channel shows boundary flows with broken chiral symmetry. Boundary flows have a constant width independent of channel dimensions which can be fit to an exponential with decay length 23 $\mu$m (dashed red line, $R^2={0.99}$) (inset). (e) Average flow for a simulated chiral active nematic also shows boundary flows with broken chiral symmetry and constant length-scale (inset). (f) Average cell orientation at the edge (red) and center (blue) of the channel showing boundary alignment with a chiral tilt. (g) Average director orientation at the edge (red) and center (blue) for a simulated chiral active nematic, again with boundary alignment with a distinct chiral tilt. Simulation units and parameter values are given in the Methods section.}
\end{figure*}

Motivated by studies of sarcoma cancer invasion in vascular networks \cite{Weigelin2012,Weigelin2016}, here we investigate collective cell migration in cultures of human fibrosarcoma cells (HT1080) confined in adhesive stripes, whose width $L_{x}$ varies in the range $300$ to $1000$\microns, surrounded by repellent coating \cite{Duclos2018a} (see Movie 1 in Supplementary Note). Although initially sparse upon the plating, cells proliferate and eventually form a confluent monolayer without visible gaps, a process which typically takes on the order of 24 h (Fig.~\ref{fig:F1}a). The spatial organization of the cellular layer is revealed by its orientational anisotropy (Fig. \ref{fig:F1}b), characterized in terms of the two-dimensional nematic tensor $\bm{Q}=S(\bm{n}\bm{n}-\mathbb{1}/2)$. Where the vector $\bm{n}$ is the nematic director and $S$ is the order parameter, describing the average orientation and degree of alignment, respectively. $\bm{Q}$ was measured from phase-contrast images with a structure tensor method (see Methods). We find a non-vanishing order parameter $\langle S\rangle=0.19 \pm 0.02$ (mean $\pm$ standard deviation for $24$ stripes), which remains constant over the measurement time, indicating that nematic order is prominent in confined fibrosarcoma cells (see Fig. S1 in Supplementary Note). This value is comparable to human bronchial epithelial cells (HBEC) under similar conditions and larger than Madin-Darby Canine Kidney  (MDCK) cell layers \cite{Blanch-Mercader2018m}. By contrast, the nematic director exhibits significant variations across the channel, resulting into a dense distribution of $+1/2$ (marked in red) and $-1/2$ (marked in blue) topological defects. 

Fig.~\ref{fig:F1}c shows a particle image velocimetry (PIV) reconstruction of the collective cellular flow. Unlike epithelial cell types, such as MDCK or HBEC \cite{Blanch-Mercader2018m}, which relax toward a jammed state after reaching confluence, in fibrosarcoma layers the root mean square speed $v_{\rm rms}=\sqrt{\langle|\bm{v}|^{2}\rangle}$ remains constant for over $25$ h, with $v_{\rm rms}=12.1\pm 0.6$\microns/h (mean $\pm$ standard deviation for $40 stripes$, see Fig. S2 in Supplementary Note). The flow is truly chaotic in the bulk, with regions of positive (red) and negative (blue) vorticity tiling the channel in an irregular and yet homogeneous fashion. Remarkably, however, an emergent structure appears at the boundary when the flow fields are averaged over the length of the channel and in time (Fig. \ref{fig:F1}d). This consists of a net tangential flow localized within a boundary layer of approximate size $23$\microns, independent on the channel width (Fig. \ref{fig:F1}d inset), and oppositely directed at the two lateral edges of the channel. Despite the channel being perfectly left-right symmetric, such an edge current is predominantly clockwise (see Fig. S3 in Supplementary Note), thus indicating a break down of chiral symmetry observable at the length scale of the entire system.

To shed light on the emergence of these chiral edge currents, we have complemented our {\em in vitro} experiments with a computational model based on the hydrodynamic theory of active nematic liquid crystals. Similar approaches have been  successfully used to account for the emergent behavior of a vast class of living systems, such as bacteria suspensions, \cite{Dunkel2013, Beer2020, Copenhagen2020, You2018} and epithelial cell monolayers \cite{Lee2011, Cochet-Escartin2014b, Banerjee2015,Blanch-Mercader2017,Tlili2018, Recho2020}. A central concept in active nemato-hydrodynamics is the so called active stress. This results from microscopic forces generated by the active sub-units along their longitudinal direction. This force can be written $\bm{F}=\pm F_{\parallel}\bm{\nu}$ , where $\bm{\nu}$ is a vector describing the orientation of individual sub-units. At the mesoscopic scale, and in the absence of microscopic chirality, these give rise to an active stress of the form $\bm{\sigma}^{\rm a}=\alpha\bm{Q}$. The phenomenological constant $\alpha$ embodies the biomechanical activity of individual sub-units and is given by $\alpha = 2a\rho F_{\parallel}$, with $a$ the cell length and $\rho$ the number density \cite{Pedley1992,Simha2002}. In the presence of chirality at the cellular scale, resulting for instance from a repositioning of the internal organelles, microscopic forces are augmented with a transverse component: i.e. $\bm{F}= \pm (F_{\parallel}\bm{\nu}+F_{\perp}\bm{\nu}^{\perp})$, with $\bm{\nu}\cdot\bm{\nu}^{\perp}=0$. This gives rise to an active stress of the form:
\begin{equation}\label{eq:active_stress}
\bm{\sigma}^{\rm a} = \alpha\bm{Q}-2\tau\bm{\epsilon}\cdot\bm{Q}\;,
\end{equation}
where $\tau = a\rho F_{\perp}$ and $\bm{\epsilon}$ is the antisymmetric Levi-Civita tensor \cite{Hoffmann2020}. With Eq. \eqref{eq:active_stress} in hand, we have numerically integrated the hydrodynamic equations of a chiral active nematic in a two-dimensional $L_{x}\times L_{y}$ rectangular domain, subject to periodic boundary conditions at $y=\pm L_{y}/2$ and stress-free boundary conditions $x=\pm L_{x}/2$ (see Methods for details). Fig.~\ref{fig:F1}e shows the average longitudinal component of the velocity field, $\langle v_y\rangle$, as a function of the distance $x$ from the channel center-line. Consistent with our experimental observations, the flow is chaotic in the bulk with vanishing time-averaged velocity, but characterized by chiral edge currents, penetrating inside the bulk by an amount independent on the channel width (Fig.~\ref{fig:F1}e inset). 

Analogously, in both our {\em in vitro} and {\em in silico} cell layers, the nematic director is randomly oriented in the bulk of the channel and parallel to the lateral edges at the boundary, with a slight tilt in the direction of the flow (Figs.~\ref{fig:F1}f,g).

\begin{figure}[t]
\centering
\includegraphics[width=\columnwidth]{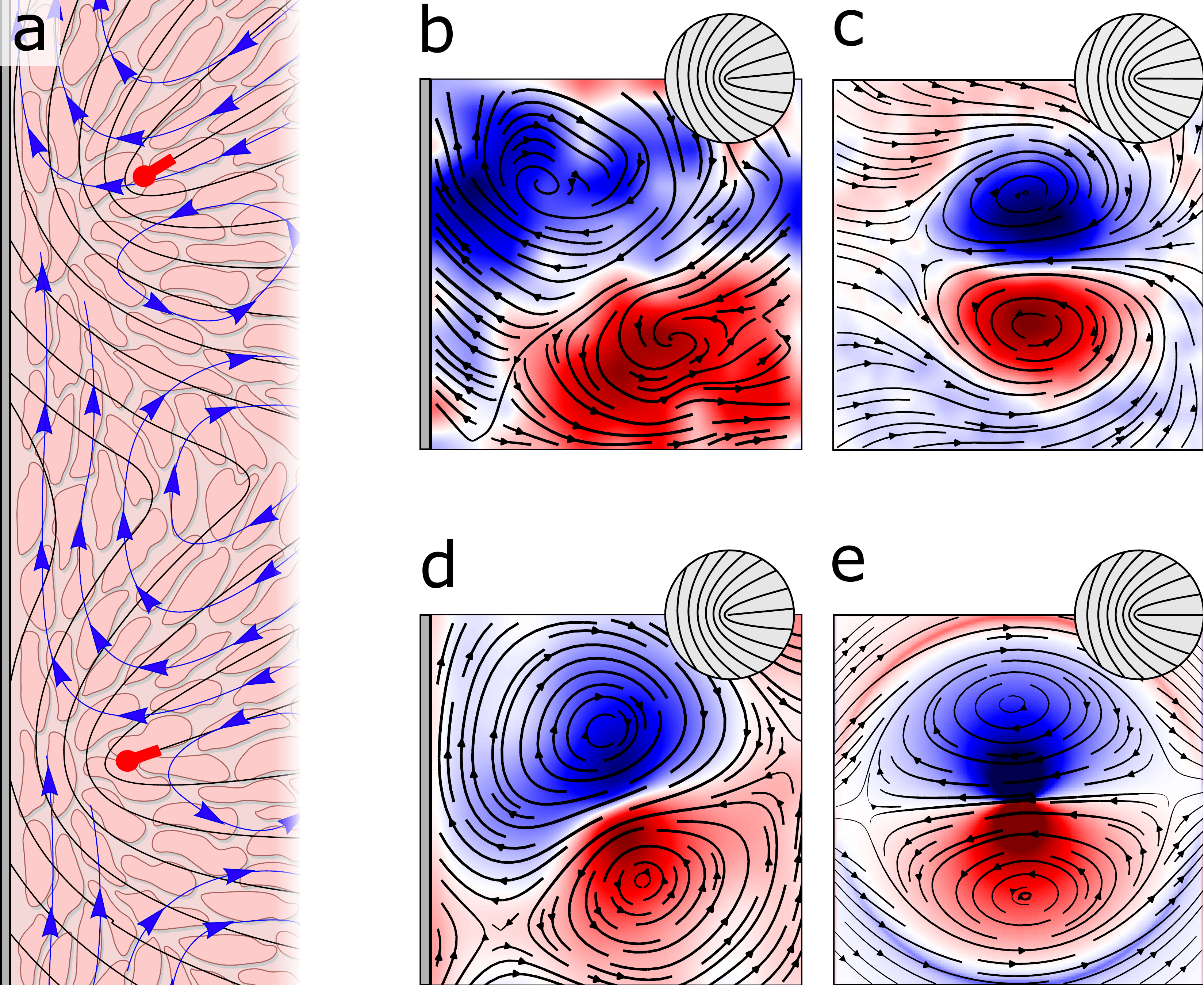}
\caption{\label{fig:F2} (a) Schematic of the hypothesised behaviour of defects close to the boundary. Motion of $+1/2$ defects and elastic interactions lead to an accumulation of defects aligned perpendicular to the boundary. The chiral activity leads to a net flow along the boundary. 
(b) Average flow around $56729$ $+1/2$ defects in a HT1080 monolayer far from the boundary with a chiral tilt of $3.4^\circ$ relative to the defect orientation (inset).
(c) Average flow around $143185$ $+1/2$ defects in a HT1080 monolayer close to the left boundary (black line). Inset shows average defect orientation close to the boundary.
(d) Average flow around $+1/2$ defects made over $10^6$ timesteps in a simulated chiral active nematic close to the left boundary (black line).
(e) Stokeslet-like flow around a chiral active $+1/2$ defect obtained from the analytical with ($\alpha=-1$, $\tau = -0.125$) oriented pointing left (inset).
Inset shows average defect orientation close to the boundary.
Panel dimensions in (b) and (c) correspond to $250~\mu$m $\times$ $250~\mu$m.}
\end{figure}

In the following we demonstrate that the edge currents observed in our experiments and simulations have a topological origin, which could be ascribed to the concerted action of $+1/2$ disclinations, orthogonally anchored to the channel walls, and the hydrodynamic flow sourcing from the chiral active stresses. In nematic cell monolayers, disclinations are point-like singularities where the orientation of the cells is undefined and around which the nematic director $\bm{n}$ rotates by $2\pi s$, with $s=\pm 1/2,\,\pm 1\ldots$ the {\em winding number}. Unlike passive nematics, where topological defects annihilate under the effect of their elastic interactions, two-dimensional active nematics can feature a steady density of $\pm 1/2$ disclinations resulting from the instability of the nematic director under the flow sourced by the active stress \cite{Giomi2015}. Although originating from the active flow, $\pm 1/2$ disclinations leave a distinct signature on the flow itself, whose local structure is almost entirely determined by the configuration of the nematic director around the defect core, via the so called {\em backflow} mechanism \cite{Giomi2013,Giomi2014}. In particular, $+1/2$ disclinations in non-chiral active nematics drive a Stokeslet-like local flow consisting of two vortices mirror-symmetrically counter-rotating about the defect longitudinal direction $\bm{p}=\nabla\cdot\bm{Q}/|\nabla\cdot\bm{Q}|$~\cite{vromans2015} (Fig.~\ref{fig:F2}). This characteristic flow pattern causes a local flow at the centre of the defect parallel with its orientation $\bm{p}$ causing $+1/2$ defects to self propel. 

Chirality affects the flow around a $+1/2$ defect by tilting its mirror-symmetry axis with respect to $\bm{p}$ by an angle $\theta_{\rm tilt}=\arctan (2\tau/\alpha)$, with $\alpha$ and $\tau$ the active stresses appearing in Eq. \eqref{eq:active_stress}, Fig.~\ref{fig:F2}b \cite{Hoffmann2020}. The latter property can be understood by noticing that the active part of the stress tensor can be expressed as $\bm{\sigma}^{\rm a}=\sqrt{\alpha^{2}+4\tau^{2}}\,\bm{R}\cdot\bm{Q}\cdot\bm{R}^{T}$, where $\bm{R}$ represents a counterclockwise rotation by an angle $\theta_{\rm tilt}/2$ and $T$ denotes transposition. 

When we analysed the average flows in the vicinity of $+1/2$ defects in HT1080 monolayers in boundary-free cell cultures the average flow field revealed a vortex-pair flow pattern, Fig.~\ref{fig:F2}c (see Methods for details). This corresponds to an extensile active stress, $\alpha<0$, with an average speed of $0.67\pm 0.05 $~$\mu {\rm m}/{\rm h}$ at the defect core (mean $\pm$ standard deviation for $56729$ defects). This average flow at the core of the defect is tilted relative to the orientation of the defect by $3.4^\circ\pm0.3$ (mean $\pm$ standard deviation for) indicating a chiral stress of the order of $\tau/\alpha = 0.03$. When the average flow field was computed for an equal-sized set of random points in boundary-free cell cultures, the vortex-pair pattern was lost, Fig.~S5 in Supplementary Note, and the average speed decreased to $0.03\pm 0.01$~$\mu {\rm m}/{\rm h}$ (mean $\pm$ standard deviation for $56729$ defects).

When the director field orients tangentially at the boundaries of a channel, elastic interactions tend to align boundary-adjacent $+1/2$ disclinations towards the boundaries, namely $\bm{p}=-\bm{N}$ with $\bm{N}$ the outer unit normal vector, Fig. \ref{fig:F2}a. This results in an enrichment of $+1/2$ defects along the edges and whose mean orientation is parallel to the outward point normal vector $\bm{N}$. Since each defect generates a vortex pair, this originates vortex street of alternating clockwise and counterclockwise vortices accumulated along the edges Fig.~\ref{fig:F2}a. Now, although representing a spectacular example of self-organization in active matter, in the absence of chirality this emergent defect structure would not give rise to an edge current due to the mirror-symmetry of the local flow generated by $+1/2$ defects. The chiral stresses, embodied in the constant $\tau$ in Eq.~\eqref{eq:active_stress}, explicitly breaks this mirror symmetry by titling individual vortex pairs with respect to the defect orientation $\bm{p}$. This tilt could push the right (or left for $\tau<0$) handed vortex of each pair closer to the boundary, leading to a boundary layer with net vorticity, thus a shear flow. Moreover, as the nematic director can be rotated by a shear flow, the latter feeds back on the cell orientation by inducing a slight but visible rotation of the defect orientation $\bm{p}$ toward the flow, thus further reinforcing the chiral edge current. This is confirmed by analysing the flow around $+1/2$ defects located close to the boundary. In both simulations and experiments, the average flow pattern around a $+1/2$ defect close to the boundary exhibited a vortex pair that was tilted with respect to boundary normal, Fig. \ref{fig:F2}d,e.

\begin{figure}[t]
\centering
\includegraphics[width=\columnwidth]{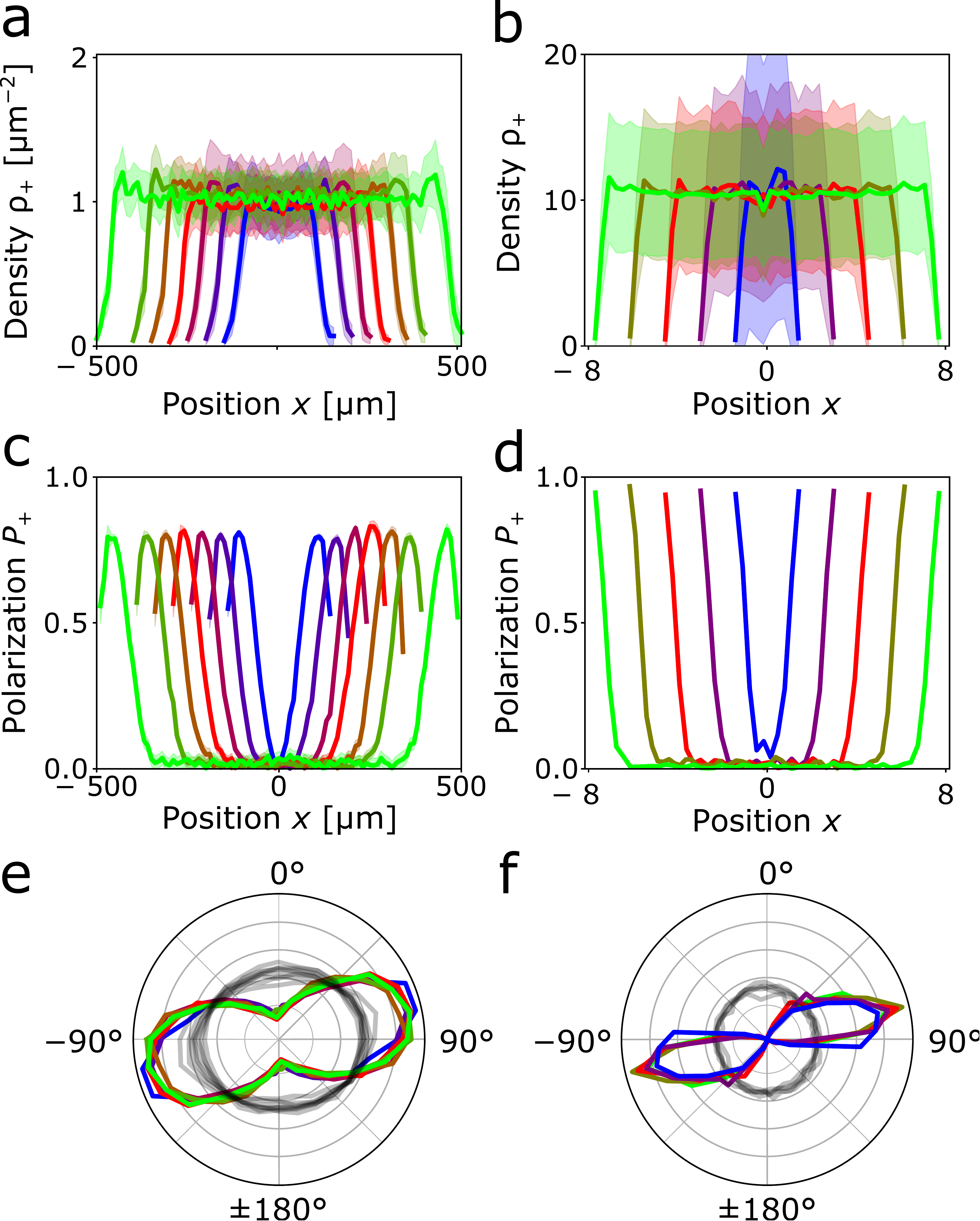}
\caption{\label{fig:F3} (a,b) Density of $+1/2$ defects in (a) monolayers of HT1080 cells and (b) simulated chiral active nematics. Defect density is normalised to the boundary-free value. (c,d) Polar order parameter of $+1/2$ defects in (a) monolayers of HT1080 cells and (b) simulated chiral active nematics, showing orientational ordering of defects close to the boundary. (e,f) Polar histogram of defect orientation close to the boundaries for (e) monolayers of HT1080 cells and (f) simulated chiral active nematics. The defects are on average aligned perpendicular to the boundary with a chiral tilt. In the bulk there is no net alignment (grey). Colour code indicates the width of the stripe as in Fig.~1. Simulation units and parameter values are given in the Methods section. }
\end{figure}

In order to further test the proposed mechanism, we compared space-dependent defect properties for stripes of varying width in both simulations and experiments, Fig.~\ref{fig:F3}. To this end, we quantify the local positive defect density as $\rho_{+}(x)= N_+(x)/\Delta xL_y$, where $N_+(x)$ is the number of $+1/2$ defects in a thin stripe of width $\Delta x$ centred at position $x$ and running the length of the channel, $L_y$. This is given by $N_+(x) = \sum_{i}\int_{x-\Delta x/2}^{x+\Delta x/2} {\rm d}x'\,\int_{-L_{y}/2}^{L_{y}/2}{\rm d}y'\,\delta(\bm{r'}-\bm{r}_{i})$, with the summation extended over all $+1/2$ defects in the system at positions $\bm{r}_{i}$. Similarly, the local positive defect polarization can be quantified $\bm{P}_{+}(x)=1/N_+(x)\sum_{i}\int_{x-\Delta x/2}^{x+\Delta x/2} {\rm d}x'\,\int_{-L_{y}/2}^{L_{y}/2}{\rm d}y'\,\bm{p}_i\,\delta(\bm{r'}-\bm{r}_{i})$. The system features a steady and spatially uniform distribution of the $+1/2$ disclinations away from the boundaries, Fig.~\ref{fig:F3}a,b. In the interior of the channel, these have no preferential orientation, as marked by the fact that the average amplitude of the polarization $P_{+}=|\bm{P}_{+}|$ vanishes, Fig.~\ref{fig:F3}c,d. By contrast, $P_{+}$ has a maximum close to the boundaries, indicating the existence of a persistent layer of co-aligned $+1/2$ defects, Fig.~\ref{fig:F3}c,d. Figs.~\ref{fig:F3}e,f show the polar histograms of the $+1/2$ defect direction $\bm{p}$ in the bulk and near the boundaries of the channel. Whereas there is no alignment in the bulk, at the boundaries we found a predominant orthogonal alignment independent of the channel width with a average chiral tilt of $16^\circ$ for HT1080 cell monolayers. This confirms the aforementioned orthogonaly anchored boundary layer of $+1/2$ defects with a chiral tilt that is independent of the channel geometry.

\begin{figure}[t]
\centering
\includegraphics[width=\columnwidth]{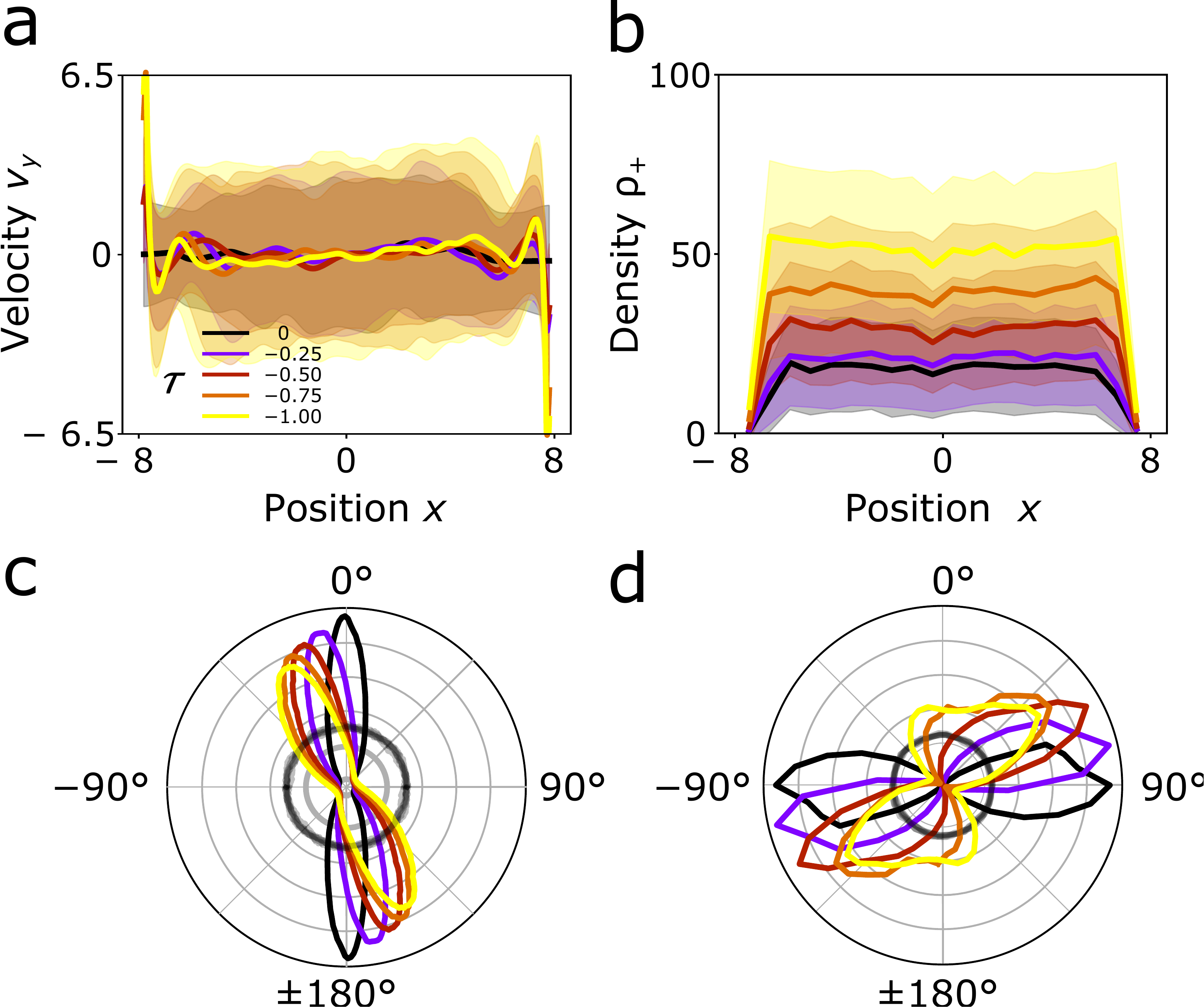}
\caption{\label{fig:F4} (a) Average flow for simulated chiral active nematics. The net chiral flow increases in magnitude for increased chiral stress. (b) The defect density increases with increased chiral stress. (c) The average tilt of the director close to the boundary increases with increased chiral stress. (d) Average $+1/2$ defect orientation close to the boundary. The magnitude of the chiral tilt increases with the chiral stress. Simulation units and parameter values are given in the Methods section. }
\end{figure}

Using numerical simulations we can explore the effect of the chiral active stress $\tau$ on the performance of the edge current. As shown in Fig. \ref{fig:F4}, both the speed $v_{y}$ of the edge current (Fig. \ref{fig:F4}a) and the density of $+1/2$ defects (Fig. \ref{fig:F4}b) increase with $\tau$. The latter, in particular, implies a reduction of the inter-defect spacing, which, in turn, causes the chiral boundary layer to become narrower and sharper. This results in a higher local shear rate, which further accentuates the rotation of the nematic director (Fig.~\ref{fig:F4}c), hence the tilt of the defect layer (Fig.~\ref{fig:F4}d).

In conclusion, when confined on adhesive stripes, HT1080 monolayers organise into a novel collective state, characterised by the existence of chiral edge currents.
Far from the boundaries, the cells exhibit disordered chaotic flows, with no net motion. The cellular flow is, however, rectified along the edges, leading to long-ranged collective cell migration, with broken chiral symmetry. Modeling the system as a chiral active nematic liquid crystal, we demonstrated that, consistently with experimental observations on stripes of various width, the edge currents are fuelled by layers of $+1/2$ topological defects, orthogonally anchored at the channel walls and acting as local sources of chiral active stress. 
Overall, our findings strengthens the idea that biological organization could take advantage of topological mechanisms \cite{Ladoux2017,Kawaguchi2017}, by demonstrating that topological defects in cell monolayers could act as organising centers for collective migration in the presence of chiral active stresses. 

\acknowledgements
It is a pleasure to thank the members of the Biology-Inspired Physics at Mesoscales (BiPMS) group. This work is partially supported by the Netherlands Organisation for Scientific Research (NWO/OCW), as part of the Vidi scheme and the Frontiers of Nanoscience program and by the ERC-CoG grant HexaTissue (L.G.).


\section{Methods}

\textbf{Cell culture.} HT1080 cells (gift from Dr Philippe Chavrier, Institut curie) were cultured in Dulbecco's modified Eagle's medium (High glucose + GlutaMAX, Gibco) supplemented with $10\%$ FBS (Sigma) and $1\%$ antibiotics solution [penicillin ($10,000$ units/mL) + streptomycin ($10$ mg/mL); Gibco] at $37$ $^\circ C$, $5\%$ $CO_2$, and $90\%$ humidity.
 
\textbf{Time-lapse microscopy.} Time-lapse multi-field experiments were performed in phase contrast on an automated inverted microscope (Olympus IX71) equipped with thermal and $CO_2$ regulations. Typical field of view (FOV) was $1.5$ mm $\times$ $1.5$ mm. The position of the measured sample and the acquisitions with a CCD camera (Retiga 4000R, QImaging) were controlled by Metamorph (Universal Imaging) software. The typical delay between two successive images of the same field was set to 15 minutes. 

\textbf{Image Processing.} Stripe were cropped from the raw images using the ImageJ public domain software \citep{imageJ}. The orientation field was obtained by computing the local structure tensor with ImageJ plugin OrientationJ  \citep{OrientationJ} within widows of $23.75~\mu$m $\times$ $23.75~\mu$m. S corresponds to the anisotropy level and the nematic director $\bm{n}$ was computed from the output angle. Visualisation of the orientation field was performed with Line Integral Convolution (LIC) under Matlab \citep{MATLAB:2018}. The velocity field in the monolayer was mapped by particle image velocimetry (PIV) analysis. Stacks of images were analysed with a custom made PIV algorithm based on the MatPIV \cite{MatPIV, Petitjean2010} software package. The window size was set to $32$ pixels = $23.75${\microns} with a $0.5$ overlap for $L > 30${\microns} and $16$ pixels = $11.9${\microns} with a $0.5$ overlap for $L=20${\microns} and $L=30${\microns}. 

\textbf{Micro-patterning technique.} Clean glass substrates were first uniformly coated with a cell-repellent layer (interpenetrated gel of acrylamide and polyethylene glycol). A photoresist mask was then structured directly on top of the layer by classical photolithography methods and air plasma was used to locally etch the protein-repellent coating through this mask. The photoresist was then removed with acetone yielding a cell repellant substrate where bare glass domains on which cells can adhere have been defined \citep{deforet2014,Duclos2018a}.

\textbf{Statistical analysis.} Statistical analysis was performed with Matlab. Experiments were performed in at least 4 replicas, each using 6 well plates with stripes of three distinct widths range (for $L=300-400${\microns} and $400-600${\microns}) or distinct widths (for $L=700${\microns}, $800${\microns}, $1000${\microns}) and plain glass slides. The number of analyzed FOVs for each width is reported in the list below. Error bars represent the SDs over all the FOVs analysed (pooling all experiments for specific stripe width $\pm5\%$). Monolayers reach confluency at least 12h after cells were seeded. The orientation and velocity were averaged over a $25$~hour period starting after confluence. List of the stripes binned by widths $L\pm5\%$ analysed in the present study and the corresponding number N of FOVs used for the orientation field and PIV analysis cells: (L($\mu$m), N) =  (300, 95); (400, 40); (500, 68); (600,47); (700, 38); (800, 43); (1000, 29); (boundary-free, 24).

\textbf{Average fields around defects.} In the following, we explain the calculation of the averaged velocity and averaged director fields over defect populations. $+1/2$ and $-1/2$ defect positions were detected by searching for the local minima of an order parameter in the window of $20~\mu$m $\times$ $20~\mu$m in the orientation map obtained by OrientationJ plugin. $+1/2$ defects and their direction was determined using the procedure explained in Ref.~\cite{vromans2015}. For each detected defect, the velocity and director maps were aligned with respect to the corresponding defect direction and then cropped over a window of size $250~\mu$m $\times$ $250~\mu$m centered at the defect core. Finally, we computed the ensemble averages over different sets of $+1/2$ defects to obtain the averaged fields. In the case of the averaged fields for random positions that are shown in Fig.~S5 in Supplementary Note, the above-mentioned procedure was the same except that the defect positions were replaced by random positions within the cell monolayer.

\textbf{Numerical simulations.} We numerically solve the hydrodynamic equations for the nematic tensor $\bm{Q}$ and the flow velocity $\bm{v}$. These are given by:
\begin{subequations}\label{eq:hydrodynamics}
\begin{gather}
(\partial_{t}+\bm{v}\cdot\nabla)\bm{Q} = \lambda S \bm{u}+\bm{Q}\cdot\bm{\omega}-\bm{\omega}\cdot\bm{Q}+\gamma^{-1}\bm{H},\\
\rho (\partial_{t}+\bm{v}\cdot\nabla)\bm{v} = \eta \nabla^{2}\bm{v}+\nabla\cdot(-P\mathbb{1}+\bm{\sigma}^{\rm e}+\bm{\sigma}^{\rm a}) -\zeta\bm{v},
\end{gather}
\end{subequations}
where $u_{ij}=(\partial_{i}v_{j}+\partial_{j}v_{i})/2$ and $\omega_{ij}=(\partial_{i}v_{j}-\partial_{j}v_{}i)/2$ are, respectively, the strain rate and vorticity tensor, $\lambda$ is the flow alignment parameter, dictating how the nematic director rotates in a shear flow, and $\gamma$ is a rotational viscosity. $P$ is the hydrostatic pressure and $\mathbb{1}$ is the identity matrix. The quantity $\bm{H}=-\delta F/\delta\bm{Q}$, with
\begin{equation}\label{eq:F}
F = \frac{1}{2}K \int {\rm d}A\, \left[|\nabla\bm{Q}|^{2}+\frac{1}{\epsilon^{2}}\tr\bm{Q}^{2}\left(\tr\bm{Q}^2-1\right)\right]\;,
\end{equation}
the Landau-de Gennes free energy of the system, is the molecular tensor driving the relaxation of the nematic tensor toward the lowest free energy configuration. $K$ is a constant with dimensions of energy, whereas $\epsilon$ is a length scale setting the typical core radius of the topological defects. In Eq. (\ref{eq:hydrodynamics}b), $\bm{\sigma}^{\rm e}=-\lambda S\bm{u}+\bm{Q}\cdot\bm{H}-\bm{H}\cdot\bm{Q}$ is the elastic stress resulting from a flow-driven distortion of the nematic director, whereas $\bm{\sigma}^{\rm a}$ is the chiral active stress given in Eq. \eqref{eq:active_stress}.
Eqns.~\eqref{eq:hydrodynamics} are numerically integrated using a finite difference scheme on a rectangular grid of size $L_x \times L_y$, subject to periodic boundary conditions at $y=\pm L_{y}/2$ and Navier boundary conditions, i.e. $\bm{\sigma}\cdot\bm{N}=-\xi\bm{v}$, with $\bm{N}$ the boundary normal and $\xi$ a drag coefficient, at $x=\pm L_{x}/2$. The aspect ratio of the channels was simulated at $L_y/L_x = 2$ and the resolution of the grid was $\Delta x = \Delta y = 0.03\,\epsilon$.
To render Eqs. (2) dimensionless we rescale length by core radius $\epsilon$, time by the nematic relaxational time scale $\tau_{\rm r}=\gamma \epsilon^{2}/K$ and stress by elastic stress $\sigma = K/\epsilon^{2}$. All quantities plotted in Figs. 1, 3 and 4 are rescaled accordingly.  In this units, we set $\lambda=0.5$, $\rho=0.1$, $\eta=0.3$, $\xi=10$, $\zeta=10^{3}$, $\alpha=-100$, $\tau=0,\,-25,\,-50,\,-57,-100$ and $L_{x}=3,\,6,\,9,\,13,\,16$. 




%

\end{document}